\documentclass[12pt]{article}

\usepackage{amssymb}
\usepackage{amsmath}

\usepackage{graphicx}

\begin{document} %


\title{Probing anomalous $\gamma\gamma\gamma\gamma$ couplings at a future muon collider}

\author{H. Amarkhail\thanks{Electronic address: hidayatamarkhail@gmail.com}
    \\
    {\small Department of Physics, Sivas Cumhuriyet University, 58140,
        Sivas, Turkey}
    \\
    {\small and Department of Physics, Kandahar University, Kandahar 3801, Afghanistan}
    \\
    S.C. \.{I}nan\thanks{Electronic address: sceminan@cumhuriyet.edu.tr}
    \\
    {\small Department of Physics, Sivas Cumhuriyet University, 58140,
        Sivas, Turkey}
    \\
    {\small and}
    \\
    A.V. Kisselev\thanks{Electronic address:
        alexandre.kisselev@ihep.ru} \\
    {\small A.A. Logunov Institute for High Energy Physics, NRC
        ``Kurchatov Institute'',}
    \\
    {\small 142281, Protvino, Russian Federation} }

\date{}

  \maketitle

\begin{abstract}
We have studied anomalous four-photon couplings in the $\mu^+\mu^-
\rightarrow \mu^+ \gamma\gamma \mu^-$ scattering at a future muon
collider. The collision energies of 3 TeV, 14 TeV, and 100 TeV are
addressed. Both differential and total cross sections versus
invariant mass of the outgoing photons are calculated. The best 95\%
C.L. exclusion bounds on anomalous couplings are obtained to be $g_1
= 2.23 \times 10^{-8}$ TeV$^{-4}$ and $g_2 = 4.22 \times 10^{-8}$
TeV$^{-4}$. They correspond to the muon collision energy of 100 TeV.
The partial-wave unitary constraints on $g_1$ and $g_2$ are
examined. We have demonstrated that the unitarity is not violated in
a region of the anomalous couplings obtained in the present paper.
\end{abstract}

\maketitle


\section{Introduction} %

In our previous papers, we examined the anomalous quartic gauge
couplings (AQGCs) of the neutral bosons in the processes
$\gamma\gamma\rightarrow\gamma\gamma$ \cite{I_K:2021_1} and
$\gamma\gamma\rightarrow\gamma Z$ \cite{I_K:2021_2} at the CLIC. It
enabled us to probe the anomalous vertices
$\gamma\gamma\gamma\gamma$ and $\gamma\gamma\gamma Z$.

The LHC experimental bounds on neutral AQGCs have been presented by
the ATLAS \cite{ATLAS:QGCs_1}-\cite{ATLAS:QGCs_2} and CMS
\cite{CMS:QGCs_1,CMS:QGCs_2} collaborations. Recently new limits on
AQGCs have been obtained by the CMS and TOTEM collaborations
\cite{CMS_TOTEM:QGCs} (see also \cite{Roland:2022}). The LHC limits
on the anomalous triple gauge couplings (ATGCs) were also given
\cite{ATLAS:TGCs}. The sensitivity of AQGCs at future hadron-hadron
colliders was investigated in \cite{Senol:2022_1,Senol:2023}.
Previously a limit on neutral AQGC was obtained at the LEP using
data on the $Z \rightarrow\gamma\gamma\gamma$ decay at the LEP
\cite{L3:Z_decay}. At present and future $e^+e^-$, $e\gamma$, and
$\gamma\gamma$ colliders AQGCs were probed in a number of papers
\cite{I_K:2021_1,I_K:2021_2},
\cite{Eboli:1994}-\cite{Gutierrez:2023}.

While in \cite{I_K:2021_1,I_K:2021_2} the $\gamma\gamma$ operation
mode of the CLIC was considered, in the present paper we investigate
AQGCs at a future high energy muon collider operating in the
$\mu^+\mu^-$ mode. A large interest in designing the high energy
muon collider is based on its capability to probe the physics beyond
the Standard Model (SM) \cite{Tikhonin:1968}-\cite{MCC:2023}. In
particular, it could provide us a determination of the Higgs
couplings with a better accuracy than what is expected at other
lepton and hadron future colliders
\cite{Barger:1997_2}-\cite{Ruhdorfer:2023}. In a number of papers
BSM interactions at the muon colliders were examined. Among them are
the SUSY particles \cite{Capdevilla:2021_1}, WIMPs
\cite{Han:2021}-\cite{Franceschini:2022}, and dark matter
\cite{Jueid:2023}, vector boson fusion \cite{Costantini:2020},
leptoquarks \cite{Asadi:2021}, lepton flavor violation
\cite{Bossi:2020}-\cite{Haghighat:2022}, vector-like leptons
\cite{Guo:2023}, heavy leptons \cite{LI:2023}-\cite{Kwok:2023}, and
heavy neutrinos \cite{Chakraborty:2022}, Majorana neutrinos
\cite{Jiang:2023}, top Yukawa couplings \cite{Chen:2022},
multi-boson processes \cite{Bredt:2022}, and physics of the muon
($g-2$) \cite{Capdevilla:2021_2}-\cite{Arakawa:2022}. In our recent
papers we have probed the warped extra dimension \cite{I_K:ED} and
axion-like particles \cite{I_K:ALP} at high energy muon colliders.
Note also several papers on searching for AQGCs
\cite{{Abbot:2022}}-\cite{Dong:2023} and ATGCs
\cite{Senol:2022_2,Spor:2022} at the muon collider.

\section{Diphoton production at muon collider} %

\begin{figure}[htb]
\begin{center}
\includegraphics[scale=0.5]{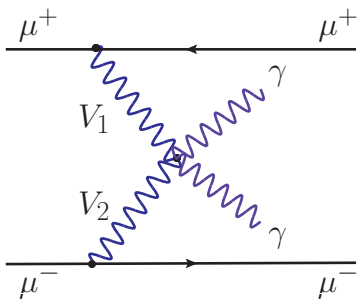}
\caption{The Feynman diagrams describing diphoton production in the
$\mu^+\mu^-$ collision via vector boson fusion.}
\label{fig:mu-VBF-mu}
\end{center}
\end{figure}

Our goal is to examine the process
\begin{equation}\label{process}
\mu^+\mu^- \rightarrow \mu^+ V_1 V_2 \mu^- \rightarrow
\mu^{-}\gamma\gamma \mu^{+} .
\end{equation}
It is shown in Fig.~\ref{fig:mu-VBF-mu}. The colliding muon beams
are assumed to be unpolarized, and we sum over the photon
polarization states. This process can be regarded as an
\emph{exclusive} process by requiring the outgoing muons to be
observable in the detector coverage
\begin{equation}\label{angle_cut}
10^\circ < \theta < 170^\circ \;,
\end{equation}
where $\theta$ is a scattering angle of the outgoing muons
\cite{Han:2023}.
%

In the equivalent photon approximation (EPA)
\cite{Weizsacker:1934}-\cite{Carimalo:1979} the polarized
distributions of photon inside unpolarized fermion beam look like
\cite{Budnev:1975}
\begin{align}\label{photon_spectrum}
f_{\gamma_\pm/f}(x, Q^2) = f_{\gamma/f}(x, Q^2) =
\frac{\alpha}{4\pi} \frac{1 + (1 - x)^2}{x} \ln\frac{Q^2}{m_\mu^2}
\;,
\end{align}
where $f = \mu^-$, $x = E_\gamma/E_\mu$ is the ratio of the photon
energy $E_\gamma$ and energy of the incoming muon $E_\mu$, $m_\mu$
is the muon mass. Note that in our case $f = \mu^-$. Because of C
and P invariance, $f_{\gamma/\bar{f}} = f_{\gamma/f}$, where
$\bar{f}$ denotes anti-fermion ($\bar{f} = \mu^+$).

For collisions of massive vector bosons ($W^\pm$ and $Z$), the
effective $W$ approximation (EWA) is usually used
\cite{Dawson:1985,Kane:1984}. It allows to treat of massive vector
bosons as partons inside the colliding beams
\cite{Cahn:1984}-\cite{Ruiz:2021}. In this approach the $Z$ boson
has the following distributions for its transverse ($\pm 1$) and
longitudinal ($0$) polarizations
\cite{Costantini:2020,Lindfors:1987,Ruiz:2021}
\begin{align}\label{Z_spectrum}
f_{Z_\pm/f}(x, Q^2) &= \frac{\alpha_Z}{4\pi} \frac{(g_V^f \mp
g_A^f)^2 + (g_V^f \pm g_A^f)^2(1 - x)^2}{x} \ln\frac{Q^2}{m_Z^2} \;,
\nonumber \\
f_{Z_0/f}(x, Q^2) &= \frac{\alpha_Z}{\pi} \frac{[(g_V^f)^2 +
(g_A^f)^2](1 - x)}{x} \;,
\end{align}
where
\begin{equation}\label{alpha_Z}
\alpha_Z = \frac{\alpha}{(\cos\theta_W \sin\theta_W)^2} \;, \ \
g_V^f = \frac{1}{2}(T_3^f)_L - Q^f \sin^2 \theta_W \;, \ \  g_A^f =
- \frac{1}{2}(T_3^f)_L  \;.
\end{equation}
Again, it is assumed that we deal with the unpolarized fermion beam.
The $Z$ boson distributions in anti-fermions are related to those in
fermions by CP relations, $f_{Z_\pm/\bar{f}} = f_{Z_\mp/f}$,
$f_{Z_0/\bar{f}} = f_{Z_0/f}$.

The cross section of the process \eqref{process} is given by
\begin{equation}\label{cs}
d\sigma = \int\limits_{\tau_{\min}}^{\tau_{\max}} \!\!d\tau
\!\!\!\int\limits_{x_{\min}}^{x_{\max}} \!\!\frac{dx}{x}
\,\sum_{V_1, V_2} \!\!f_{V_1/\mu^+}(x, Q^2) f_{V_2/\mu^-}(\tau/x,
Q^2) \,d\hat{\sigma} (V_1V_2\rightarrow \gamma\gamma) \;,
\end{equation}
where $V_1$ and $V_2$ run over $\gamma, Z_+, Z_-, Z_0$,
\begin{equation}\label{y_z_limits}
x_{\max} = 1 - \frac{m_\mu}{E_\mu} \;, \ \tau_{\max} = \left( 1 -
\frac{m_\mu}{E_\mu} \right)^{\!2} , \ x_{\min} = \tau/x_{\max} \;, \
\tau_{\min} = \frac{p_\bot^2}{E_\mu^2} \;,
\end{equation}
and $p_\bot$ is the transverse momenta of the outgoing photons. The
boson distributions inside the muon beam, $f_{\gamma/{\mu^\pm}}(x,
Q^2)$, $f_{Z_\pm/{\mu^\pm}}(x, Q^2)$, and $f_{Z_0/{\mu^\pm}}(x,
Q^2)$ are given by eqs.~\eqref{photon_spectrum}, \eqref{Z_spectrum}.
We take $Q^2 = \hat{s}$, where $\sqrt{\hat{s}} = 2E_\mu \sqrt{\tau}$
is the invariant energy of the VBF subprocess
$V_1V_2\rightarrow\gamma\gamma$. The differential cross section in
\eqref{cs} is a sum of squared helicity amplitudes,
\begin{equation}\label{subprocess_cs}
\frac{d\hat{\sigma}}{d\Omega}(V_1V_2\rightarrow \gamma\gamma) =
\frac{1}{64\pi^2 \hat{s}} \sum_{\lambda_3,\lambda_4}
\!\!|M_{\lambda_1\lambda_2\lambda_3\lambda_4}|^2 ,
\end{equation}
where $\lambda_1, \lambda_2$ are helicity of the incoming bosons,
and $\lambda_3, \lambda_4$ are helicities of the outgoing photons.
Remind that we consider the unpolarized diphoton final state. In its
turn, each of the helicity amplitudes
$M_{\lambda_1\lambda_2\lambda_3\lambda_4}$ in \eqref{subprocess_cs}
is a sum of the anomaly and SM terms,
\begin{equation}\label{anom+SM}
M_{\lambda_1\lambda_2\lambda_3\lambda_4} =
M_{\lambda_1\lambda_2\lambda_3\lambda_4}^{\mathrm{anom}} +
M_{\lambda_1\lambda_2\lambda_3\lambda_4}^{\mathrm{SM}} \;.
\end{equation}

In what follows, to obtain exclusion bounds for the AQGCs, we assume
that there are no anomalous vertices $Z\gamma\gamma\gamma$ and
$ZZ\gamma\gamma$. It means that only the
$\gamma\gamma\rightarrow\gamma\gamma$ subprocess can contribute to
$d\hat{\sigma} (V_1V_2\rightarrow \gamma\gamma)$ (in other words,
one has to put $V_1 = V_2 = \gamma$ in eq.~\eqref{cs}). Otherwise,
one may obtain only bound on some combination of anomalous couplings
corresponding to the $\gamma\gamma\gamma\gamma$,
$Z\gamma\gamma\gamma$, and $ZZ\gamma\gamma$ vertices. Since the
distribution of the photon in the unpolarized muon does not depend
on photon's helicity \eqref{photon_spectrum}, we have to average
4-photon amplitude in \eqref{subprocess_cs} over helicity states
$\lambda_1, \lambda_2$.

To calculate the anomalous helicity amplitudes we use an effective
Lagrangian which contributes to anomalous quartic boson vertices. It
can be found, for instance, in \cite{Gupta:2012}
\begin{align}\label{Lagrangian_QNGCs}
\mathcal{L}_{\mathrm{anom}} & = \frac{c_8}{\Lambda^4} B_{\rho\sigma}
B^{\rho\sigma} B_{\mu\nu} B^{\mu\nu} + \frac{c_9}{\Lambda^4}
W_{\rho\sigma}^a W^{a\rho\sigma} W_{\mu\nu}^b W^{b\mu\nu} +
\frac{c_{10}}{\Lambda^4} W_{\rho\sigma}^a W^{b\rho\sigma}
W_{\mu\nu}^a W^{b\mu\nu} \nonumber \\
&+ \frac{c_{11}}{\Lambda^4} B_{\rho\sigma} B^{\rho\sigma}
W_{\mu\nu}^a W^{a\mu\nu} + \frac{c_{13}}{\Lambda^4} B_{\rho\sigma}
B^{\sigma\nu} B_{\nu\mu} B^{\mu\rho} + \frac{c_{14}}{\Lambda^4}
W_{\rho\sigma}^a W^{a\sigma\nu} W_{\nu\mu}^b W^{b\mu\rho} \nonumber \\
&+ \frac{c_{15}}{\Lambda^4} W_{\rho\sigma}^a W^{b\sigma\nu}
W_{\nu\mu}^a W^{b\mu\rho} + \frac{c_{16}}{\Lambda^4} B_{\rho\sigma}
B^{\sigma\nu} W_{\nu\mu}^a W^{a\mu\rho}.
\end{align}
In a broken phase this Lagrangian is expressed in terms of the
physical bosons. We are interested in the part of the effective
Lagrangian that describes the anomalous $\gamma\gamma\gamma\gamma$
interaction. It is given by
\begin{equation}\label{contact_Lagrangian_gamma}
\mathcal{L}_{\gamma\gamma\gamma\gamma} = g_1
F_{\mu\nu}F^{\mu\nu}F_{\rho\sigma} F^{\rho\sigma} + g_2
F_{\mu\nu}F^{\mu\rho}F_{\rho\sigma} F^{\sigma\nu} ,
\end{equation}
where $F_{\mu\nu} = \partial_\mu A_\nu - \partial_\nu A_\mu$, and
the couplings $g_1$ and $g_2$ have dimension TeV$^{-4}$. The
couplings $g_1, g_2$ are given by the following linear combinations
of $c_i$ ($i = 8, \ldots, 16$) \cite{Gupta:2012}
\begin{align}\label{gi_vs_ci}
g_1 &= [c^4_w c_8 + s^4_w c_9 + c^2_w s^2_w (c_{10} + c_{11})]
\Lambda^{-4} \;,
\nonumber \\
g_2 &= [c^4_w c_13 + s^4_w c_14 + c^2_w s^2_w (c_{15} + c_{16})]
\Lambda^{-4} \;.
\end{align}
Instead of the second operator in \eqref{contact_Lagrangian_gamma}
the $F_{\mu\nu}\tilde{F}^{\mu\nu}F_{\rho\sigma}
\tilde{F}^{\rho\sigma}$ operator are sometimes encountered, where
$F^{\mu\nu} = (1/2)\epsilon^{\mu\nu\alpha\beta}F_{\alpha\beta}$. The
corresponding Lagrangians can be transformed into each other (up to
a full derivative) using equations of motion.

As we see, two operators in \eqref{contact_Lagrangian_gamma} can
arise from the $SU(2)\times U(1)_Y$ effective Lagrangian which
contains products of four field strengths of the hypercharge gauge
field $B_\mu$ and $SU(2)$ gauge field $W_\mu$. However, we may not
appeal to the effective Lagrangian \eqref{Lagrangian_QNGCs} at all,
but just to demand that our operators must be dimension-8 Lorentz
invariant CP-even operators of four photon fields. Then we should
come to the same operators \eqref{contact_Lagrangian_gamma} (or to a
physically equivalent set of two photon operators, see comments
after \eqref{gi_vs_ci}). In other words, anomalous interactions in
the broken phase \eqref{contact_Lagrangian_gamma} can be considered
as independent.

In a number of papers bounds on couplings $c_i$ (often denoted as
$f_{T,i}$) in EF Lagrangian \eqref{Lagrangian_QNGCs} were searched
for. While obtaining bounds on any coupling, the other couplings
were set to zero (see, for instance,
\cite{Koksal:2023,Gutierrez:2023}).%
\footnote{If in the unbroken phase \eqref{Lagrangian_QNGCs} all
couplings $c_i$ except one are assumed to be zero, then AQGCs are
defined by one and the same quantity. For instance, if $c_8 \neq 0$,
and $c_i = 0$ for $i=9, \ldots, 16$, then
$4g^{\gamma\gamma\gamma\gamma} = -(c_w/s_w)g^{Z\gamma\gamma\gamma} =
-(c_w/s_w)^3g^{ZZZ\gamma}
 = 4(c_w/s_w)^4g^{ZZZZ}$ \cite{Gupta:2012}.}
We prefer to examine couplings of \emph{physical} operators (like
couplings $g_1, g_2$ in the effective Lagrangian
\eqref{contact_Lagrangian_gamma}). As mentioned above, we will study
four-photon anomalous interaction assuming that there are no quartic
interactions of the photons with one or two $Z$ bosons.%
\footnote{Analogously, AQGCs for the $Z\gamma\gamma\gamma$ and
$ZZ\gamma\gamma$ vertices can be examined.}

The explicit expressions for
$M_{\lambda_1\lambda_2\lambda_3\lambda_4}^{\mathrm{anom}}$ are
collected in Appendix. As for the SM contribution to the cross
section, all the terms in \eqref{cs} must be taken into account
($V_1, V_2 = \gamma, Z$). For our numerical calculations we will use
explicit expressions for the SM helicity amplitudes obtained for the
processes $\gamma\gamma\rightarrow\gamma\gamma$
\cite{Jikia:1994}-\cite{Atag:2010}, $\gamma\gamma\rightarrow\gamma
Z$ \cite{Gounaris:1999_3}, and $\gamma\gamma\rightarrow ZZ$
\cite{Gounaris:2000}. The $W$-loop contribution to the helicity
amplitudes is the leading contribution. Moreover, in the region
\begin{equation}\label{energy_region}
s \sim |t| \sim |u| \gg (m_W^2, m_Z^2)
\end{equation}
the dominant amplitudes are $M_{++++}$ and $M_{+-+-}$ (for instance,
see \cite{I_K:ampl}), as well as helicity amplitudes related to them
by Bose statistics, crossing symmetry, and parity or time-inversion
invariance. Also note that in the high energy region
\eqref{energy_region} the relative size of the helicity amplitudes
for $\gamma\gamma\rightarrow\gamma\gamma$ and
$\gamma\gamma\rightarrow ZZ$ collisions is given by the relation
$M(\gamma\gamma\rightarrow\gamma\gamma) = (\tan \theta_W)^4
M(\gamma\gamma\rightarrow ZZ)$ \cite{Berger:2005}.

\begin{figure}[htb]
\begin{center}
\includegraphics[scale=0.52]{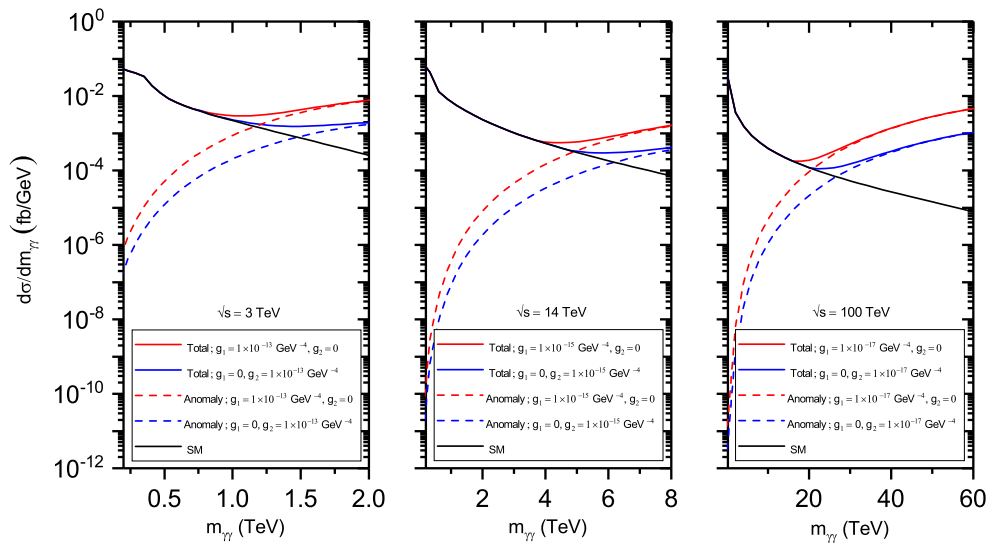}
\caption{The differential cross sections for the  $\mu^+\mu^-
\rightarrow \mu^+ \gamma\gamma \mu^-$ scattering at the future muon
collider versus diphoton invariant mass $m_{\gamma\gamma}$. The
center-of-mass energy is equal to 3 TeV, 14 TeV, and 100 TeV.}
\label{fig:MD4GAMMAS}
\end{center}
\end{figure}

The differential cross sections for different values of the
collision energy $\sqrt{s}$ are presented in
Fig.~\ref{fig:MD4GAMMAS} as functions of the diphoton invariant mass
$m_{\gamma\gamma}$. We have used the cut on the rapidity of the
outgoing photons, $|\eta| < 2.5$. The curves for two sets of the
anomalous couplings $g_1$, $g_2$ are shown. The invariant mass
$m_{\gamma\gamma}$ of the diphoton system is expected to be close to
the collision energy. But the center mass energy could be
smeared by ISR/meanstrahlung effects.%
\footnote{The muon collider has a reduced ISR, contrary to $e^+e^-$
collider. It means that the tail of the beam energy spread is not
large.}
To suppress this SM background, we will use the cut
$m_{\gamma\gamma} < m_{\gamma\gamma, \max} = 0.9 \sqrt{s}$. Our cut
could not reduce the total cross section noticeably, since for all
three values of the collision energy the differential cross section
decreases rapidly as $m_{\gamma\gamma}$ approaches $\sqrt{s}$, see
Fig.~\ref{fig:MD4GAMMAS}. The anomalous part of the differential
cross section becomes to dominate the SM part at large
$m_{\gamma\gamma}$. This effect is more pronounced for $\sqrt{s} =
14$ TeV, and especially for $\sqrt{s} = 100$ TeV.
%

\begin{figure}[htb]
\begin{center}
\includegraphics[scale=0.52]{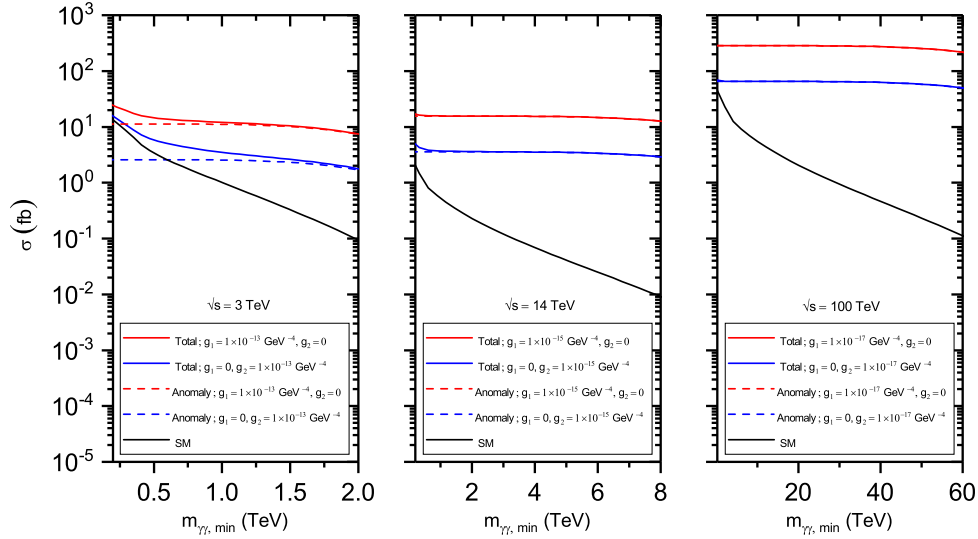}
\caption{The total cross sections $\sigma(m_{\gamma\gamma,\min} <
m_{\gamma\gamma} < m_{\gamma\gamma,\max})$ for the $\mu^+\mu^-
\rightarrow \mu^+ \gamma\gamma \mu^-$ scattering at the future muon
collider versus minimal value of the diphoton invariant mass
$m_{\gamma\gamma}$.}
\label{fig:MC4GAMMAS}
\end{center}
\end{figure}

In Fig.~\ref{fig:MC4GAMMAS} the total cross section versus minimal
value of the parameter $m_{\gamma\gamma}$ is shown. If $\sqrt{s} =
3$ TeV the anomaly part of the total cross section becomes larger
than the SM contribution for $m_{\gamma\gamma, \min}
> 0.5$ TeV. But for 14 TeV, and especially for 100 TeV, the anomaly
cross section dominates the SM one almost in the full region of
$m_{\gamma\gamma, \min}$, being several orders of magnitude more at
large $m_{\gamma\gamma, \min}$. Therefore, to get better sensitivity
bounds for the anomalous couplings, we have to use larger values of
$m_{\gamma\gamma,\min}$ (see below).
%

\begin{figure}[htb]
\begin{center}
\includegraphics[scale=0.45]{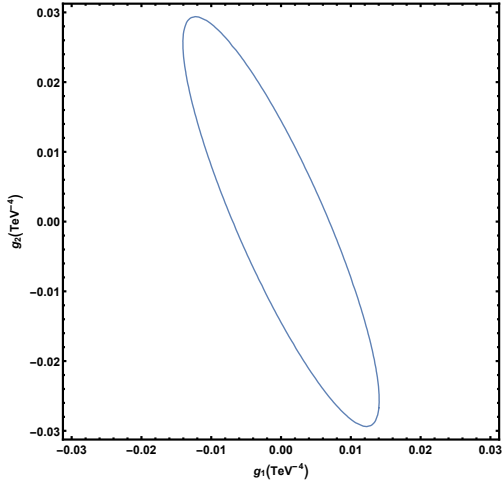}
\caption{The 95\% C.L. exclusion region for the couplings $g_1, g_2$
in unpolarized $\mu^+\mu^- \rightarrow \mu^+ \gamma\gamma \mu^-$
scattering at the muon collider. The inner region of the ellipse is
inaccessible. The collision energy is $\sqrt{s} = 3$ TeV, the
integrated luminosity is $L = 1$ ab$^{-1}$.}
\label{fig:SSE1500}
\end{center}
\end{figure}

\begin{figure}[htb]
\begin{center}
\includegraphics[scale=0.5]{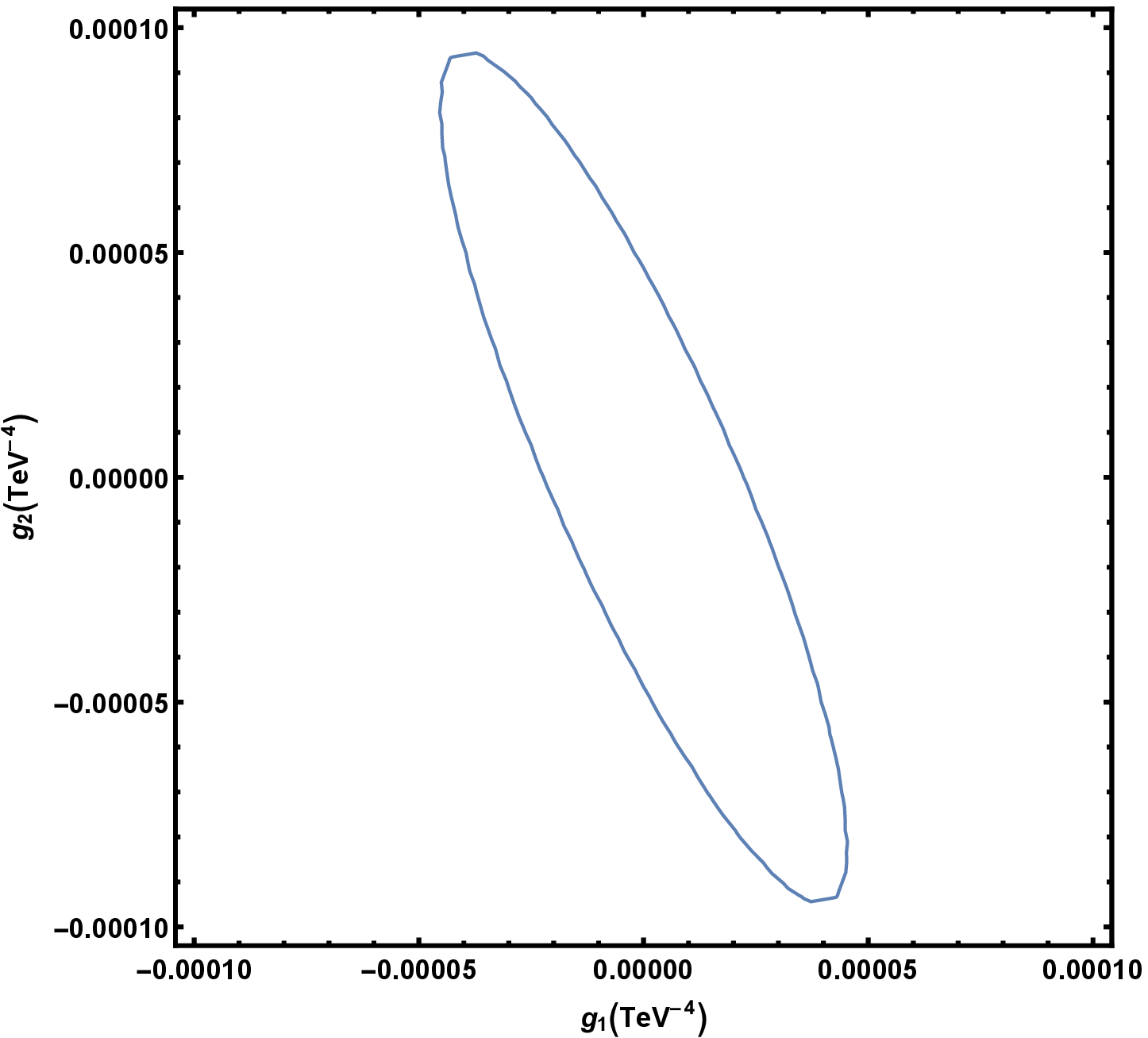}
\caption{The same as in Fig.~\ref{fig:SSE1500}, but for $\sqrt{s} =
14$ TeV and $L = 20$ ab$^{-1}$.}
\label{fig:SSE7000}
\end{center}
\end{figure}

\begin{figure}[htb]
\begin{center}
\includegraphics[scale=0.5]{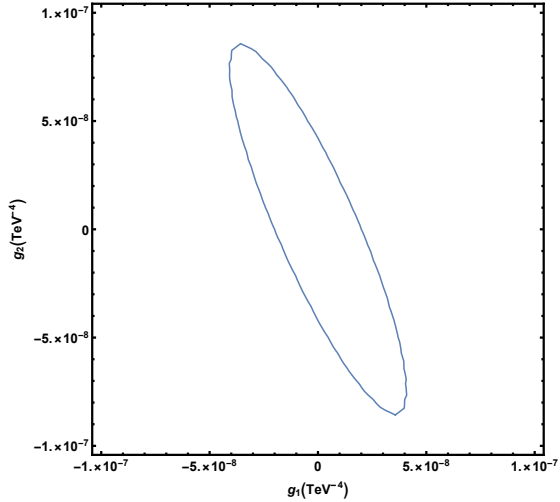}
\caption{The same as in Fig.~\ref{fig:SSE1500}, but for $\sqrt{s} =
100$ TeV and $L = 1000$ ab$^{-1}$.}
\label{fig:SSE50000}
\end{center}
\end{figure}

To obtain exclusion regions, we apply the following formula for the
statistical significance $SS$ \cite{SS}
\begin{equation}\label{SS_def}
SS = \sqrt{2[(S - B\,\ln(1 + S/B)]} \;,
\end{equation}
where $S$ is the number of signal events and $B$ is the number of
background events. We define the regions $SS \leqslant 1.645$ as the
regions that can be excluded at the 95\% C.L. To reduce the SM
background, we take additional cuts on the invariant mass of the
outgoing photons: i) $m_{\gamma\gamma} > 0.5$ TeV for $\sqrt{s} = 3$
TeV; ii) $m_{\gamma\gamma} > 3$ TeV for $\sqrt{s} = 14$ TeV; iii)
$m_{\gamma\gamma} > 20$ TeV for $\sqrt{s} = 100$ TeV. The 95\% C.L.
exclusion regions for the anomalous couplings $g_1$ and $g_2$ are
shown in Figs.~\ref{fig:SSE1500}-\ref{fig:SSE50000}. The expected
integrated luminosities are taken from \cite{MCWG:2019,Han:2021}.
Note that the total cross section is proportional to the coupling
combination $48g_1^2 + 40g_1g_2 + 11g_2^2$, if we take the helicity
amplitudes given in Appendix \cite{I_K:2021_1,Fichet:2015}. It means
that the exclusion regions are ellipses in the $(g_1 - g_2)$ plane,
and that the cross section is approximately twice more sensitive to
the coupling $g_1$ than to $g_2$.

%
%

\section{Unitarity limits on anomalous couplings} %

To examine bounds imposed by partial-wave unitarity, we will use the
partial-wave expansion of the helicity amplitude in the
center-of-mass system \cite{Jacob:2000},
\begin{align}\label{helicity_ampl_expansion}
M_{\lambda_1\lambda_2\lambda_3\lambda_4}(s, \theta, \varphi) &=
16\pi \sum_J (2J + 1) \sqrt{(1 + \delta_{\lambda_1\lambda_2})(1 +
\delta_{\lambda_3\lambda_4})}
\nonumber \\
&\times \,e^{i(\lambda - \mu)\phi} \,d^J_{\lambda\mu}(\theta)
\,T^J_{\lambda_1\lambda_2\lambda_3\lambda_4}(s) \;,
\end{align}
where $\lambda = \lambda_1 - \lambda_2$, $\mu = \lambda_3 -
\lambda_4$, $\theta(\phi)$ is the polar (azimuth) scattering angle,
and $d^J_{\lambda\mu}(\theta)$ is the Wigner (small) $d$-function
\cite{Wigner}. It obeys the relation $d^J_{\lambda\mu}(-z) = (-1)^{J
- \lambda}d^J_{\mu -\lambda}(z)$, where $z = \cos\theta$. If we
choose the plane $(x - z)$ as a scattering plane, then $\phi = 0$ in
\eqref{helicity_ampl_expansion}. In what follows, we will need only
\begin{equation}\label{d-function_00}
d^J_{00}(z) = P_J(z) \;,
\end{equation}
where $P_J(z)$ is the Legendre polynomial, and \cite{I_K:2021_2}
\begin{equation}\label{d-function_22}
d^J_{2 \pm2}(z) = (\pm 1)^J \!\left( \frac{1 \pm z}{2}\right)^{\!2}
\!{}_2F_1 \!\left( 2-J, J+3; 1; \frac{1 \mp z}{2} \right) ,
\end{equation}
where $_2F_1(a,b;c;x)$ is the hypergeometric function
\cite{Bateman_vol_1}.

Partial-wave unitarity requires that
\begin{equation}\label{parity_wave_unitarity}
\left| T^J_{\lambda_1\lambda_2\lambda_3\lambda_4}(s) \right| \leq 1
\;.
\end{equation}
Using orthogonality of the $d$-functions,
\begin{equation}\label{d_func_orthogonality}
\int\limits_{-1}^1 d^{J}_{\lambda\lambda'}(z)
\,d^{J'}_{\lambda\lambda'}(z) \,dz = \frac{2}{2J + 1} \,\delta_{JJ'}
\;,
\end{equation}
we find from \eqref{helicity_ampl_expansion} that the partial-wave
amplitude is given by
\begin{align}\label{parity wave_func}
T^J_{\lambda_1\lambda_2\lambda_3\lambda_4}(s) &= \frac{1}{32\pi}
\frac{1}{\sqrt{(1 + \delta_{\lambda_1\lambda_2})(1 +
\delta_{\lambda_3\lambda_4})}} \int\limits_{-1}^1 \!\!
M_{\lambda_1\lambda_2\lambda_3\lambda_4}(s, z)
\,d^{J}_{\lambda\mu}(z) \,dz \;.
\end{align}
The anomalous helicity amplitudes
$M_{\lambda_1\lambda_2\lambda_3\lambda_4}$ are collected in
Appendix.

\textbf{1.} Helicity amplitude $M_{++++}$. Then $\lambda_1 =
\lambda_2 = \lambda_3 = \lambda_4 = 1$, and $\lambda = \mu = 0$.
Using explicit expression for $M_{++++}$
\eqref{independent_helicity_ampl}, we find that the partial-wave
amplitude with $J=0$ is the only non-zero helicity amplitude,
\begin{equation}\label{T++++}
T^J_{++++}(s) = \frac{(4g_1 + 3g_2)s^2}{128\pi} \!\int\limits_{-1}^1
\!P_J(z) \,dz = \frac{(4g_1 + 3g_2)s^2}{64\pi} \delta_{J0} \;.
\end{equation}
We find from \eqref{parity_wave_unitarity}, \eqref{T++++} that
\begin{equation}\label{T++++_unitarity}
|4g_1 + 3g_2| \leq 64\pi s^{-2} \;.
\end{equation}

\textbf{2.} Helicity amplitude $M_{++--}$. In such a case,
$\lambda_1 = \lambda_2 = -\lambda_3 = -\lambda_4 = 1$, and $\lambda
= \mu = 0$. It follows from \eqref{independent_helicity_ampl} that
$M_{++--} = (4g_1 + g_2)s^2(3 + z^2)/4$. Then we obtain
\begin{align}\label{T++--}
T^J_{++--}(s) &= \frac{(4g_1 + g_2)s^2}{256\pi} \int\limits_{-1}^1
\!P_J(z) (3 + z^2) \,dz
\nonumber \\
&= \frac{(4g_1 + g_2)s^2}{256\pi} \!\left[ \frac{20}{3}\delta_{J0} +
\frac{4}{15}\delta_{J2} \right] .
\end{align}
Thus, non-trivial amplitudes are $T^0_{++--}(s)$ and
$T^2_{++--}(s)$. It is $T^0_{++--}(s)$ that results in more
stringent bound on the anomalous couplings, and we get
\begin{equation}\label{T++--_unitarity}
|4g_1 + g_2| \leq (192\pi/5) s^{-2} \;.
\end{equation}

\textbf{3.} Helicity amplitude $M_{+-+-}$. Then $\lambda_1 =
-\lambda_2 = \lambda_3 = -\lambda_4 = 1$, and $\lambda = \mu = 2$.
Using equation \eqref{d-function_22} and taking into account that
$M_{+-+-} = (4g_1 + 3g_2)s^2[(1 + z)/2]^2/2$, we get
\begin{align}\label{T+-+-}
T^J_{+-+-}(s) &= \frac{(4g_1 + 3g_2)s^2}{64\pi} \int\limits_{-1}^1
\left( \frac{1 + z}{2}\right)^{\!2} \!d^{J}_{22}(z) \,dz =
\frac{(4g_1 + 3g_2)s^2}{32\pi} \int\limits_0^1 (1 - x)^4
\nonumber \\
&\times {}_2F_1( 2-J, J+3; 1; x) dx = \frac{3(4g_1 +
3g_2)s^2}{4\pi\Gamma(4+J)\Gamma(3-J)} \;,
\end{align}
where $\Gamma(x)$ is the gamma function. To derive \eqref{T+-+-},
eqs.~2.21.1.5 and 2.21.1.6 in \cite{Prudnikov_v3} were applied. As
one can see from \eqref{T+-+-}, only three partial-waves amplitudes,
$T^0_{+-+-}(s)$, $T^1_{+-+-}(s)$, and $T^2_{+-+-}(s)$, are non-zero.
The strongest unitary bound on anomalous couplings comes from
$T^0_{+-+-}(s) = (4g_1 + 3g_2)s^2/(16\pi)$. It looks like
\begin{equation}\label{T+-+-_unitarity}
|4g_1 + 3g_2| \leq 16\pi s^{-2} \;.
\end{equation}

\textbf{4.} Helicity amplitude $M_{+--+}$. Since $\lambda_1 =
-\lambda_2 = -\lambda_3 = \lambda_4 = 1$, we get $\lambda = 2, \,\mu
= -2$. After accounting for eqs.~\eqref{parity wave_func},
\eqref{dependent_helicity_ampl}, and \eqref{d-function_22}, it can
be shown that we come again to inequality \eqref{T+-+-_unitarity}.

If the couplings $g_1$, $g_2$ have the \emph{same} sign, then we
obtain the following unitary bounds
\begin{equation}\label{bounds_same_sign}
|g_1| < 4\pi s^{-2} \;, \quad |g_2| < (16\pi/3) s^{-2} \;.
\end{equation}
Now let us assume that the couplings $g_1$, $g_2$ have the
\emph{opposite} sign. It is sufficient to consider the case when
$g_1 > 0$, $g_2 < 0$. There are three possibilities:

\textbf{1.} $4g_1 > 3|g_2|$. We get from \eqref{T++--_unitarity},
\eqref{T+-+-_unitarity} that
\begin{equation}\label{bounds_opposite_sign_1}
3|g_2| < 4g_1 \leq \min \left( 3|g_2| + 16\pi s^{-2}, |g_2| +
(192\pi/5) s^{-2} \right) .
\end{equation}
It follows from \eqref{bounds_opposite_sign_1} that $g_1 < (72\pi/5)
s^{-2}$ and $|g_2| < (96\pi/5) s^{-2}$.

\textbf{2.} $3|g_2| \geq 4g_1 \geq |g_2|$. Then
\begin{equation}\label{bounds_opposite_sign_2}
|g_2| \leq 4g_1 \leq 3|g_2| \leq (708\pi/5) s^{-2} \;.
\end{equation}
It gives us the upper bounds $g_1 \leq (177\pi/5) s^{-2}$ and $|g_2|
\leq (236\pi/5) s^{-2}$.

\textbf{3.} $4g_1 < |g_2|$. We obtain
\begin{equation}\label{bounds_opposite_sign_3}
3|g_2| - 16\pi s^{-2} \leq  4g_1 < |g_2| \;.
\end{equation}
From \eqref{bounds_opposite_sign_3} we come to the upper limits $g_1
< 2\pi s^{-2}$ and $|g_2| < 8\pi s^{-2}$.

As one can see from the inequalities shown above, the most stringent
upper bound on $g_1$ looks like
\begin{equation}\label{bounds_g1}
g_1 < 2\pi s^{-2} \;.
\end{equation}
So, the unitarity demands that $|g_1|$ must be less than $7.8 \times
10^{-2}$ TeV$^{-4}$, $1.64 \times 10^{-4}$ TeV$^{-4}$, and $6.28
\times 10^{-8}$ TeV$^{-4}$, for 3 TeV, 14 TeV, and 100 TeV,
respectively. The best bound on $g_2$ is achieved when $g_1 = 0$,
\begin{equation}\label{bounds_g2}
|g_2| \leq (16\pi/3) s^{-2} \;.
\end{equation}
Correspondingly, the unitary upper limits for $|g_2|$ are equal to
$0.21$ TeV$^{-4}$, $4.36 \times 10^{-4}$ TeV$^{-4}$, and $1.68
\times 10^{-7}$ TeV$^{-4}$, for 3 TeV, 14 TeV, and 100 TeV,
respectively. Comparing obtained unitary bounds with our exclusion
regions presented in Figs.~\ref{fig:SSE1500}-\ref{fig:SSE50000}, we
conclude that the unitary in not violated in our region of AQGCs.
Let us underline that for numerical calculations of the unitary
bounds, in all the above equations, we put $s$ to be equal to the
collision energy of the process $\mu^+\mu^- \rightarrow \mu^+
\gamma\gamma \mu^-$. Actually, in these equations, one has to use
the squared invariant energy of the
$\gamma\gamma\rightarrow\gamma\gamma$ subprocess $\hat{s}=sx_1x_2$
instead of $s$. As a result, effective unitary bounds are even
weaker than those given above.

\section{Conclusions} %

In the present paper we have studied the anomalous quartic couplings
of the $\gamma\gamma\gamma\gamma$ vertex in the unpolarized
$\mu^+\mu^- \rightarrow \mu^+ \gamma\gamma \mu^-$ scattering at the
future muon collider. We require the scattering angles of the
outgoing muons to be in the region  $10^\circ \div 170^\circ$
\cite{Han:2023}. To derive the bounds on the four-photon couplings,
we have assumed that anomalous couplings of the vertices
$Z\gamma\gamma\gamma$ and $ZZ\gamma\gamma$ are zero. The collision
energies of 3 TeV, 14 TeV, and 100 TeV are examined.

Using analytical expressions for the helicity amplitudes for the
light-by-light scattering, we have calculated both differential and
total cross sections. To reduce the SM background, we have imposed
the cuts: $m_{\gamma\gamma} > 0.5$ TeV, 3 TeV, and 20 TeV for
$\sqrt{s} = 3$ TeV, 14 TeV, and 100 TeV, respectively. Additionally
we demand the invariant mass of the outgoing photons to satisfy the
inequality $m_{\gamma\gamma} < 0.9\sqrt{s}$. Finally, we use the cut
on the rapidity of the final photons, $|\eta| < 2.5$. The 95\% C.L.
exclusion regions for the anomalous four-photon couplings $g_1$ and
$g_2$ are calculated. For $\sqrt{s} = 3$ TeV the best bounds
appeared to be $g_1 = 6.9 \times 10^{-3}$ TeV$^{-4}$ (for $g_2 = 0$)
and $g_2 = 1.41 \times 10^{-2}$ TeV$^{-4}$ (for $g_1 = 0$).
Correspondingly, for $\sqrt{s} = 14$ TeV we have obtained the limits
$g_1 = 2.2 \times 10^{-5}$ TeV$^{-4}$, $g_2 = 4.6 \times 10^{-5}$
TeV$^{-4}$. Finally, for $\sqrt{s} = 100$ TeV the most stringent
bounds on anomalous couplings are equal to $g_1 = 2.23 \times
10^{-8}$ TeV$^{-4}$ and $g_2 = 4.22 \times 10^{-8}$ TeV$^{-4}$.
Thus, new scales of the order of 14 TeV and 82 TeV can be probed at
the future muon collider with the collision energy of 14 TeV and 100
TeV, respectively.

The partial-wave unitary constraints on the anomalous couplings are
examined in detail. We have demonstrated that the unitarity is not
violated in the region of these values of couplings $g_1$, $g_2$.

The AQGCs can originate from one-loop contribution with a new
charged particle inside the loop. Then one expect that $g_i \sim
\alpha^2/\Lambda^4$, where $\Lambda$ is a mass scale of new physics,
and our constraint $g_2 = 4.2 \times 10^{-8}$ TeV$^{-4}$ translates
to $\Lambda \simeq 6$ TeV. However, there could be quite another
origin of the AQGCs coming from $s$-channel diagram with a new
neutral mediator. Suppose that its coupling to the SM fields is
equal to $1/f$. Then we expect that $g_i \sim 1/(f^2\Lambda^2)$.
This new particle can be, for instance, a KK graviton or radion in
the Randall-Sundrum (RS) model with one extra dimension and warped
metric \cite{Randall:1999}. Taking $1/f \sim 1$ TeV$^{-1}$ (as in
the RS model), we obtain from $g_2 = 4.2 \times 10^{-8}$ TeV$^{-4}$
that $\Lambda \simeq 500$ TeV. Thus, the use of EFT and our cuts are
justified even for the 100 TeV muon collider.

The sensitivity to the $\gamma\gamma\gamma\gamma$ anomalous
couplings in photon collisions at the HL-LHC were estimated to be
$g_1(g_2) < 1 \times 10^{-2}$ TeV$^{-4}$
\cite{Fichet:2015,Fichet:2017}. As one can see, it is comparable
with the sensitivity of the 3 TeV muon collider. The sensitivity of
the 100 TeV FCC-hh with integrated luminosity of 3 ab$^{-1}$ is
expected to be $g_1(g_2) < 2 \times 10^{-4}$ TeV$^{-4}$
\cite{Fichet:2017}, while our bounds on $g_1$ and $g_2$ for the 100
TeV muon collider are \emph{four} orders of magnitude stronger. It
demonstrates once again an advantage of the muon collider over
FCC-hh collider in searching for BSM physics.

Recently the CMS and TOTEM collaborations have presented a first
search for the diphoton production at high mass with tagged protons
at 13 TeV \cite{CMS-TOTEM:diphoton}. Interpreting the experimental
results in an effective dimension-8 extension of the SM, the
following limits at 95\% C.L. have been obtained: $|g_1| < 2.9
\times 10^{-1}$ TeV$^{-4}$ ($g_2 = 0$), $|g_2| < 6.0 \times 10^{-1}$
TeV$^{-4}$ ($g_1 = 0$) \cite{CMS-TOTEM:diphoton}.



\setcounter{equation}{0}
\renewcommand{\theequation}{A.\arabic{equation}}

\section*{Appendix. Anomalous helicity amplitudes}

Here we consider helicity amplitudes for the process
$\gamma\gamma\rightarrow\gamma\gamma$. Bose-Einstein statistics and
parity invariance demand that there exist six independent helicity
amplitudes $M_{\lambda_1\lambda_2\lambda_3\lambda_4}$ with
$\lambda_1 = +1$. Only three of them are nonzero,
\begin{align}\label{independent_helicity_ampl}
M_{++++}(s,t,u) &= \frac{(4g_1 + 3g_2)}{2} \,s^2 \;,
\nonumber \\
M_{++--}(s,t,u) &=  \frac{(4g_1 + g_2)}{2} \,(s^2 + t^2 + u^2)\;,
\nonumber \\
M_{+-+-}(s,t,u) &=  \frac{(4g_1 + 3g_2)}{2} \,u^2 \;,
\end{align}
where $s$, $t$, $u$ are Mandelstam variables. Other three
independent helicity amplitudes are equal to zero,
\begin{equation}\label{zero_helicity_ampl}
M_{+++-} = M_{++-+} = M_{+-++} = 0 \;.
\end{equation}
We also have the crossing relations
\begin{align}\label{dependent_helicity_ampl}
M_{+--+}(s,t,u) &= M_{+-+-}(s,u,t) = \frac{(4g_1 + 3g_2)}{2} \,t^2
\;,
\nonumber \\
M_{+---}(s,t,u) &= M_{+-++}(s,u,t) = 0 \;.
\end{align}
Equations
\eqref{independent_helicity_ampl}-\eqref{dependent_helicity_ampl}
represent eight helicity amplitudes with $\lambda_1 = +1$. The
helicity amplitudes with $\lambda_1 = -1$ can be obtained from them
by the use of the parity relation,
\begin{equation}\label{parity_relations}
M_{-\lambda_2\lambda_3\lambda_4}(s,t,u) =
M_{+-\lambda_2-\lambda_3-\lambda_4}(s,t,u) \;.
\end{equation}


\section*{Data availability statement}

This work has no associated data.





\begin{thebibliography}{999}
%
\bibitem{I_K:2021_1}
S.C.~\.{I}nan and A.V.~Kisselev, \emph{Probing anomalous quartic
$\gamma\gamma\gamma\gamma$ couplings in light-by-light collisions at
the CLIC}, Eur. Phys. J. C \textbf{81}, 664 (2021)
[arXiv:2009.09712].
%
\bibitem{I_K:2021_2}
S.C.~\.{I}nan and A.V.~Kisselev, \emph{Probing anomalous
$\gamma\gamma\gamma Z$ couplings through $\gamma Z$ production in
$\gamma\gamma$ collisions at the CLIC}, JHEP \textbf{10}, 121 (2021)
[arXiv:2108.04478].
%
\bibitem{Fichet:2015}
S.~Fichet, G.~von~Gersdorff, B.~Lenzi, Ch.~Royond, and M.~Saimpertd,
\emph{Light-by-light scattering with intact protons at the LHC: from
Standard model to new physics}, JHEP \textbf{02}, 165 (2015)
[arXiv:1411.6629].
%
\bibitem{Braun:2008}
H.~Braun \emph{et al.} (CLIC Study Team), \emph{CLIC 2008
parameters}, CERN-OPEN-2008-021, CLIC-NOTE-764.
%
\bibitem{Boland:2016}
M.J.~Boland \emph{et al.} (CLIC and CLICdp Collaborations),
\emph{Updated baseline for a staged Compact Linear Collider},
CERN-2016-004 [arXiv:1608.07537].
%
\bibitem{Chapon:2009}
E.~Chapon, O.~Kepka and C.~Royon, \emph{Probing $WW\gamma\gamma$ and
$ZZ\gamma\gamma$ quartic anomalous couplings with 10 pb$^{-1}$ at
the LHC}, arXiv:0908.1061.
%
\bibitem{Chapon:2010}
E.~Chapon, C.~Royon and O.~Kepka, \emph{Anomalous quartic
$WW\gamma\gamma$, $ZZ\gamma\gamma$, and trilinear $WW\gamma$
couplings in two-photon processes at high luminosity at the LHC},
Phys. Rev. D \textbf{81}, 074003 (2010).
%
\bibitem{Sahin:2012}
\.{I}.~\c{S}ahin and B.~\c{S}ahin, \emph{Anomalous quartic
$ZZ\gamma\gamma$ couplings in gamma-proton collision at the LHC},
Phys. Rev. D \textbf{86}, 115001 (2012).
%
\bibitem{Senol:2014}
A.~Senol, \emph{Anomalous quartic $WW\gamma\gamma$ and
$ZZ\gamma\gamma$ couplings in $\gamma p$ collision at the LHC}, Int.
J. Mod. Phys. A \textbf{29}, 1450148 (2014).
%
\bibitem{Eboli:2004}
O.J.P.~\'{E}boli, M.C.~Gonzalez-Garcia and S.M.~Lietti,
\emph{Bosonic quartic couplings at CERN LHC}, Phys. Rev. D
\textbf{69}, 095005 (2004).
%
\bibitem{ATLAS:QGCs_1}
G.~Aad \emph{et al.} (ATLAS Collaboration), \emph{Measurements of
$Z\gamma$ and $Z\gamma\gamma$ production in $pp$ collisions at
$\sqrt{s} = 8$ TeV with the ATLAS detector}, Phys. Rev. D
\textbf{93}, 112002 (2006).
%
\bibitem{ATLAS:Z_decay}
G.~Aad et al. (ATLAS Collaboration), \emph{Search for new phenomena
in events with at least three photons collected in $pp$ collisions
at $\sqrt{s} = 8$ TeV with the ATLAS detector}, Eur. Phys. J. C
\textbf{76}, 210 (2016) [arXiv:1509.05051].
%
\bibitem{ATLAS:QGCs_2}
ATLAS Collaboration, \emph{Measurement of electroweak
$Z\nu\bar{\nu}jj$ production and limits on anomalous quartic gauge
couplings in pp collisions at $\sqrt{s} =13$ TeV with the ATLAS
detector}, arXiv:2208.12741.
%
\bibitem{CMS:QGCs_1}
A.M.~Sirunyuan \emph{et al.} (CMS Collaboration), \emph{Measurement
of the cross section for electroweak production of a $Z$ boson, a
photon and two jets in proton-proton collisions at $\sqrt{s} = 13$
TeV and constraints on anomalous quartic couplins}, JHEP
\textbf{06}, 076 (2020) [arXiv:2002.09902].
%
\bibitem{CMS:QGCs_2}
CMS Collaboration, \emph{Measurements of the $pp \rightarrow
W^\pm\gamma\gamma$ and $pp \rightarrow Z\gamma\gamma$ cross sections
at $\sqrt{s} = 13$ TeV and limits on anomalous quartic gauge
couplings}, JHEP \textbf{10}, 174 (2021) [arXiv:2105.12780].
%
\bibitem{CMS_TOTEM:QGCs}
The CMS and TOTEM Collaborations, \emph{Search for high-mass
exclusive $\gamma\gamma\rightarrow WW$ and $\gamma\gamma\rightarrow
ZZ$ production in proton-proton collisions at $\sqrt{s} = 13$ TeV},
JHEP \textbf{07}, 229 (2023) [arXiv:2211.16320].
%
\bibitem{Roland:2022}
C.~Roland, \emph{Triple and quartic gauge boson couplings}, Talk at
the 41st International Symposium on Physics in Collision (on behalf
of the ATLAS and CMS sollaborations), 5-9 September, 2022, Tbilisi,
Georgia.
%
\bibitem{ATLAS:TGCs}
M.~Aaboud \emph{et al.} (ATLAS Collaboration), \emph{Measurement of
the $Z\gamma\rightarrow \nu\bar{\nu}\gamma $ production cross
section in $pp$ collisions at $\sqrt{s} = 13$ TeV with the ATLAS
detector and limits on anomalous triple gauge-boson couplings}, JHEP
\textbf{12}, 010 (2018).
%
\bibitem{L3:Z_decay}
M.~Acciarri et al. (L3 Collaboration), \emph{Search for anomalous
$Z\rightarrow\gamma\gamma\gamma$ events at LEP}, Phys. Lett. B
\textbf{345}, 609 (1995).
%
\bibitem{Senol:2022_1}
A.~Senol, O.~Karadeniz, K.Y.~Oyulmaz, C.~Helveci, and H.~ Denizli,
\emph{Sensitivity of anomalous quartic gauge couplings via
$Z\gamma\gamma$ production at future hadron-hadron colliders}, Nucl.
Phys. B \textbf{980}, 115851 (2022) [arXiv:2109.12572].
%
\bibitem{Senol:2023}
A.~Senol, H.~Denizli, and C.~Helveci, \emph{Sensitivity of anomalous
quartic gauge couplings via tri-photon production at FCC-hh}, Nucl.
Phys. B \textbf{998}, 116387 (2024) [arXiv:2303.14805].
%
\bibitem{Eboli:1994}
O.J.P.~\'{E}boli, M.C.~Gonzalez-Garcia and S.F.~Novaes,\emph{Quartic
anomalous couplings in $e\gamma$ colliders}, Nucl. Phys. B
\textbf{411}, 381 (1994) [arXiv:hep-ph/9306306].
%
\bibitem{Eboli:1995}
O.J.P.~\'{E}boli, M.B.~Magro, P.G.~Mercadante, and S.F.~Novaes,
\emph{Quartic anomalous couplings in $\gamma\gamma$ colliders},
Phys. Rev. D \textbf{52}, 15 (1995) [arXiv:hep-ph/9503432].
%
\bibitem{Stirling:2020}
W.J.~Stirling and A.~Werthenbach,\emph{Anomalous quartic couplings
in $W^+ W^- \gamma$, $Z^0 Z^0 \gamma$ and $Z^0 \gamma\gamma$
production at present and future $e^+ e^-$ colliders}, Eur. Phys. J.
C \textbf{14}, 103 (2000) [arXiv:hep-ph/9903315].
%
\bibitem{Atag:2007}
S.~Ata\u{g} and \.{I}.~\c{S}ahin, {Anomalous quartic
$WW\gamma\gamma$ and $ZZ\gamma\gamma$ couplings in $e\gamma$
collision with initial beams and final state polarizations}, Phys.
Rev. D \textbf{75}, 073003 (2007) [arXiv:hep-ph/0703201].
%
\bibitem{Eboli:2001_2}
O.J.P.~\'{E}boli and J. K. Mizukoshi, \emph{Probing anomalous
quartic couplings in $e\gamma$ and $\gamma\gamma$ colliders}, Phys.
Rev. D \textbf{64}, 075011 (2001) [arXiv:hep-ph/0105238].
%
\bibitem{Sahin:2009}
\.{I}.~\c{S}ahin, \emph{Anomalous quartic $WW\gamma\gamma$ and
$WWZ\gamma$ couplings through $W^+W^-Z$ production in $\gamma\gamma$
collisions}, J. Phys. G: Nucl. Part. Phys. \textbf{36}, 075007
(2009) [arXiv:0807.4777].
%
\bibitem{Koksal:2014}
M.~K\"{o}ksal, \emph{Anomalous quartic $ZZ\gamma\gamma$ couplings at
the CLIC}, Eur. Phys. J. Plus \textbf{130}, 75 (2015)
[arXiv:1402.3773].
%
\bibitem{Koksal:2016}
M.~K\"{o}ksal, V.~Ari and A.~Senol, \emph{Search for anomalous
quartic $ZZ\gamma\gamma$ couplings in photon-photon collisioms},
Adv. High Energy Phys. \textbf{2016}, 8672391 (2016)
[arXiv:1606.04433].
%
\bibitem{Gutierrez:2014}
A.~Guti\'{e}rrez-Rodr\'{i}guez, C.G.~Honorato, J.~Monta\~{n}o and
M.A.~P\'{e}rez, \emph{Limits on the quartic couplings
$Z\gamma\gamma\gamma$ and $ZZ\gamma\gamma$ from $e^+e^-$ colliders},
Phys. Rev. D \textbf{89}, 034003 (2014) [arXiv:1304.7410].
%
\bibitem{Koksal:2023}
M.~K\"{o}ksal, \emph{Search for the anomalous quartic gauge
couplings through $Z\gamma$ production at $e^-e^+$ colliders}, J.
Phys. G: Nucl. Part. Phys. \textbf{51}, 015001 (2023)
[arXiv:2306.11894].
%
\bibitem{Gutierrez:2023}
A.~Guti\'{e}rrez-Rodr\'{i}guez, E.~Gurkanli, M.~K\"{o}ksal, V.~Ari,
and M.A.~Hern\'{a}ndez-Ru\'{i}z,\emph{Expected sensitivity on the
anomalous quartic neutral gauge couplings in $\gamma\gamma$
collisions at the CLIC}, EPJ Plus \textbf{138}, 1130 (2023)
\textbf{[arXiv:2308.05210]}.
%
\bibitem{Tikhonin:1968}
F.F.~ Tikhonin, \emph{On the effects of the clashing $\mu$-meson
beams} (in Russian), JINR Report P2-4120, Dubna, 1968; \emph{On the
effects at colliding mu-meson beams}, arXiv:0805.3961.
%
\bibitem{Budker:1969}
G.I.~Budker, \emph{Accelerators and colliding beams}. In Proceedings
of the 7th International Conference on High-Energy Accelerators.
(HEACC 1969), 27 August -- 2 September 1969, Yerevan, USSR. Vol.
\textbf{1}, pp. 33-39 (1970).
%
\bibitem{Skrinsky:1981}
A.N.~Skrinsky and V.V.~Parkhomchuk, \emph{Cooling methods for beams
of charged particles} (in Russian), Sov. J. Part. Nucl. \textbf{12},
223 (1981) [Fiz. Elem. Chast. Atom. Yadra \textbf{12}, 557 (1981)].
%
\bibitem{Bartosik:2020}
N.~Bartosik, \emph{Detector and physics performance at a muon
collider}, JINST \textbf{15} P05001 (2020) [arXiv:2001.04431].
%
\bibitem{Collamati:2021}
F.~Collamati \emph{et al.}, \emph{Advanced assessment of
beam-induced background at a muon collider}, JINST \textbf{16},
P11009 (2021) [arXiv:2105.09116].
%
\bibitem{Blondel:1999}
A.~Blondel, J.R.~Ellis, and B.~Autin, \emph{Prospective study of
muon storage rings at CERN}. In CERN Yellow Reports: Monographs.
CERN: Geneva, Switzerland, 1999.
%
\bibitem{Buttazzo:2021}
D.~Buttazzo, R.~Franceschini, and A.~Wulzer, \emph{Two paths towards
precision at a very high energy lepton collider}, JHEP \textbf{05},
219 (2021) [arXiv:2012.11555].
%
\bibitem{Long:2021}
K.R.~Long, D.~Lucchesi, M.A.~Palmer, N.~Pastrone, D.~Schulte, and
V.~Shiltsev, \emph{Muon colliders to expand frontiers of particle
physics}, Nature Phys. \textbf{17}, 289 (2021) [arXiv:2007.15684].
%
\bibitem{MCWG:2019}
J.P.~Delahaye \emph{et al.} (Muon Collider Working Group),
\emph{Muon colliders}, arXiv:1901.06150
%
\bibitem{MCFR:2022}
K.M.~Black \emph{et al.}, \emph{Muon Collider Forum Report},
arXiv:2209.01318.
%
\bibitem{MCC:2023}
C.~Accettura \emph{et al.} (Muon Collider Collaboration),
\emph{Towards a muon collider}, arXiv:2303.08533.
%
\bibitem{Barger:1997_2}
V.~Barger, M.S.~Berger, J.F.~Gunion, and T.~Han, \emph{Higgs boson
physics in the $s$-channel at $\mu^+\mu^-$ colliders}, Phys. Rep.
\textbf{286}, 1 (1997) [arXiv:hep-ph/9602415].
%
\bibitem{Chiesa:2020}
M.~Chiesa, \emph{Measuring the quartic Higgs self-coupling at a
multi-TeV muon collider}, JHEP \textbf{09}, 098 (2020)
[arXiv:2003.13628].
%
\bibitem{Franceschini:2021}
R.~Franceschini and M.~Greco, \emph{Higgs and BSM physics at the
future muon collider}, Symmetry \textbf{13}, 851 (2021)
[arXiv:2104.05770].
%
\bibitem{Bandyopadhyay:2021}
P.~Bandyopadhyay and A.~Costantini, \emph{Obscure Higgs boson at
colliders}, Phys. Rev. D \textbf{103}, 015025 (2021)
[arXiv:2010.02597].
%
\bibitem{Han:2021_1}
T.~Han, D.~Liu, I.~Low, and X.~Wang, \emph{Electroweak couplings of
the Higgs boson at a multi-TeV muon collider}, Phys. Rev. D
\textbf{103}, 013002 (2021) [arXiv:2008.12204].
%
\bibitem{Han:2021_2}
T.~Han, S.~Li, S.~Su, W.~Su, and Y.~Wu, \emph{Heavy Higgs bosons in
2HDM at a muon collider}, Phys. Rev. D \textbf{104}, 055029 (2021)
[arXiv:2102.08386].
%
\bibitem{Han:2021_3}
T.~Han \emph{et al.}, \emph{Precision test of the muon-Higgs
coupling at a high-energy muon collider}, JHEP \textbf{12}, 162
(2021) [arXiv:2108.05362].
%
\bibitem{Costantini:2021}
A.~Costantini, \emph{(New) physics at a multi-TeV $\mu$ collider},
arXiv:2111.02507.
%
\bibitem{Forslund:2022}
M.~Forslund and P.~Meade, \emph{High precision Higgs from high
energy muon colliders}, JHEP \textbf{08}, 175 (2022)
[arXiv:2203.09425].
%
\bibitem{Han:2022}
T.~Han, S.~Li, S.~Su, W.~Su, and Y.~Wu, \emph{BSM Higgs production
at a muon collider}, arXiv:2205.11730.
%
\bibitem{Ruhdorfer:2023}
M.~Ruhdorfer, E.~Salvioni, and A.~Wulzer, \emph{Invisible Higgs from
forward muons at a muon collider}, Phys. Rev. D \textbf{107}, 095038
(2023) [arXiv:2303.14202].
\bibitem{Capdevilla:2021_1}
R.~Capdevilla, F.~Meloni, R.~Simoniellod, and J.~Zurita,
\emph{Hunting wino and higgsino dark matter at the muon collider
with disappearing tracks}, JHEP \textbf{06}, 133 (2021)
[arXiv:2102.11292].
%
\bibitem{Han:2021}
T.~Han, Z.~Liu, L.-T.~Wang, and X.~Wang, \emph{WIMPs at high energy
muon colliders}, Phys. Rev. D \textbf{103}, 075004 (2021)
[arXiv:2009.11287].
%
\bibitem{Black:2022}
K.~Black \emph{et al.}, \emph{Prospects for heavy WIMP dark matter
searches at muon colliders}, arXiv:2205.10404.
%
\bibitem{Franceschini:2022}
R.~Franceschini and X.~Zhao, \emph{Going all the way in the search
for WIMP dark matter at the muon collider through precision
measurements}, arXiv:2212.11900.
%
\bibitem{Jueid:2023}
A.~Jueid and S.~Nasri, \emph{Lepton portal dark matter at muon
colliders: total rates and generic features for phenomenologically
viable scenarios}, Phys. Rev. D \textbf{107}, 115027 (2023)
[arXiv:2301.12524].
%
\bibitem{Costantini:2020}
A.~Costantini \emph{et al.}, \emph{Vector boson fusion at multi-TeV
muon colliders}, JHEP \textbf{10}, 080 (2020) [arXiv:2005.10289].
%
\bibitem{Asadi:2021}
P.~Asadi, R.~Capdevilla, C.~Cesarotti, and S.~Homiller,
\emph{Searching for leptoquarks at future muon colliders}, JHEP
\textbf{10}, 182 (2021) [arXiv:2104.05720].
%
\bibitem{Bossi:2020}
F.~Bossi and P.~Ciafaloni, \emph{Lepton flavor violation at
muon-electron colliders}, JHEP \textbf{10}, 033 (2020)
[arXiv:2003.03997].
%
\bibitem{Homiller:2022}
S.~Homiller, Q.~Lu, and M.~Reece, \emph{Complementary signals of
lepton flavor violation at a high-energy muon collider}, JHEP
\textbf{07}, 036 (2022) [arXiv:2203.08825].
%
\bibitem{Haghighat:2022}
G.~Haghighat, M.M.~Najafabadi, \emph{Search for
lepton-flavor-violating decays of the tau lepton at a future muon
collider}, arXiv:2204.04433.
%
\bibitem{Guo:2023}
Q.~Guo, L.~Gao, Y.~Mao, and Q.~Li, \emph{Search for vector-like
leptons at a muon collider}, arXiv:2204.01885.
%
\bibitem{LI:2023}
P.~Li, Z.~Liu, and K.-F.~Lyu, \emph{Heavy neutral leptons at muon
colliders}, JHEP \textbf{03}, 231 (2023) [arXiv:2301.07117].
%
\bibitem{Mekala:2023}
K.~Mekala, J.~Reuter, and A.F.~Zarnecki, \emph{Optimal search reach
for heavy neutral leptons at a muon collider}, Phys. Lett. B
\textbf{841}, 137945 (2023) [arXiv:2301.02602].
%
\bibitem{Kwok:2023}
T.H.~Kwok, L.~Li, T.~Liu, and A.~Rock, \emph{Searching for heavy
neutral leptons at a future muon collider}, arXiv:2301.05177.
%
\bibitem{Chakraborty:2022}
I.~Chakraborty, H.~Roy, and T.~Srivastava, \emph{Searches for heavy
neutrinos at multi-TeV muon collider: a resonant leptogenesis
perspective}, Eur. Phys. J. C \textbf{23}, 180 (2023)
[arXiv:2206.07037].
%
\bibitem{Jiang:2023}
R.~Jiang \emph{et al.}, \emph{Searching for Majorana neutrinos at a
same-sign muon collider}, arXiv:2304.04483.
%
\bibitem{Chen:2022}
M.~Chen and D.~Liu, \emph{Top Yukawa coupling at the muon collider},
arXiv:2212.11067.
%
\bibitem{Bredt:2022}
P.~Bredt, W.~Kilian, J.~Reuter, and P.~Stienemeier, \emph{NLO
electroweak corrections to multi-boson processes at a muon
collider}, JHEP \textbf{12}, 138 (2023) [arXiv:2208.09438].
%
\bibitem{Capdevilla:2021_2}
R.~Capdevilla, D.~Curtin, Y.~Kahn, and G.~Krnjaic, \emph{Discovering
the physics of $(g-2)_\mu$ at future muon colliders}, Phys. Rev. D
\textbf{103}, 075028 (2021) [arXiv:2006.16277].
%
\bibitem{Yin:2022}
W.~Yin and M.~Yamaguchi, \emph{Muon $g-2$ at multi-TeV muon
collider}, Phys. Rev. D \textbf{106}, 033007 (2022)
[arXiv:2012.03928].
%
\bibitem{Arakawa:2022}
J.~Arakawa, A.~Rajaraman, T.~Sui, and T.M.P.~Tait, \emph{Probing
muon $g-2$ at a future muon collider}, arXiv:2208.14464.
%
\bibitem{I_K:ED}
S.C.~Inan, and A.V.~Kisselev, \emph{Probe of a Randall-Sundrum-like
model from muon pair production at high energy muon collider},
arXiv:2301.08585.
%
\bibitem{I_K:ALP}
S.C.~Inan, and A.V.~Kisselev, \emph{Probe of axion-like particles in
vector boson scattering at a muon collider},  J. Phys. G: Nucl.
Part. Phys. \textbf{50} 105002 (2023) [arXiv:2207.03325].
%
\bibitem{Abbot:2022}
B.~Abbot \emph{et al.}, \emph{Anomalous quartic gauge couplings at a
muon collider}, in Proceedings of the US Community Study on the
Future of Particle Physics (Snowmass 2021), arXiv:2203.08135.
%
\bibitem{Yang:2022_1}
J.-C.~Yang, X.-Y.~Han, Z.-B.~Qin, T.~Li, and Y.-C.~Guo,
\emph{Measuring the anomalous quartic gauge couplings in the $W^+W^-
\rightarrow W^+W^- $ process at muon collider using artificial
neural networks}, JHEP \textbf{09}, 074 (2022) [arXiv:2204.10034].
%
\bibitem{Yang:2022_2}
J.-C.~Yang, Z.-B.~Qing, X.-Y.~Han, Y.-C.~Guo, and T.~Li,
\emph{Tri-photon at muon collider: a new process to probe the
anomalous quartic gauge couplings}, JHEP \textbf{07}, 053 (2022)
[arXiv:2204.08195].
%
\bibitem{Dong:2023}
Y.-F.~Dong, Y.-C.~Mao, and J.-C.~Yanga, \emph{Searching for
anomalous quartic gauge couplings at muon colliders using principle
component analysis}, Eur. Phys. J. C \textbf{83}, 555 (2023)
[arXiv:2304.0150].
%
\bibitem{Senol:2022_2}
A.~Senol, S.~Spor, E.~Gurkanli, V.~Cetinkaya, H.~Denizli, and
M.~K\"{o}ksal, \emph{Model-independent study on the anomalous
$ZZ\gamma$ and $Z\gamma\gamma$ couplings at the future muon
collider}, Eur. Phys. J. Plus \textbf{137}, 1354 (2022).
%
\bibitem{Spor:2022}
S.~Spor, \emph{Probe of the anomalous neutral triple gauge couplings
in photon-induced collision at future muon colliders}, Nucl. Phys. B
\textbf{991}, 116198 (2023) [arXiv:2207.11585].
%
\bibitem{Han:2023}
T.~Han, T.~Li, and X.~Wang, \emph{Axion-like particles at high
energy muon colliders -- A white paper for Snowmass 2021},
arXiv:2203.05484.
%
\bibitem{Weizsacker:1934}
C.F.~Weizs\"{a}cker, \emph{Ausstrahlung bei St\"{o}{\ss}en sehr
schneller Elektronen}, Z. Phys. \textbf{88}, 612 (1934).
%
\bibitem{Williams:1934}
E.J.~Williams, \emph{Nature of the high energy particles of
penetrating radiation and status of ionization and radiation
formulae}, Phys. Rev. \textbf{45}, 729 (1934).
%
\bibitem{Brodsky:1971}
S.J.~Brodsky, \emph{Two-photon mechanism of particle production by
high-energy colliding beams}, Phys. Rev. D \textbf{4}, 1532 (1971).
%
\bibitem{Terazawa:1973}
H.~Terazawa, \emph{Two-photon processes for particle production at
high energies}, Rev. Mod. Phys. \textbf{45}, 615 (1973).
%
\bibitem{Budnev:1975}
V.M.~Budnev, I.F.~Ginzburg, G.V.~Meledin, and V.G.~Serbo, \emph{The
two photon particle production mechanism. Physical problems.
Applications. Equivalent photon approximation}, Phys. Rep.
\textbf{15}, 181 (1975).
%
\bibitem{Carimalo:1979}
C.~Carimalo, P.~Kessler, and J.~Parisi, \emph{Validity of the
equivalent-photon approximation for virtual photon-photon
collisions}, Phys. Rev. D \textbf{20}, 1057 (1979).
%
\bibitem{Dawson:1985}
S.~Dawson, \emph{The effective $W$ approximation}, Nucl. Phys. B
\textbf{249}, 42 (1985).
%
\bibitem{Kane:1984}
G.L.~Kane, W.W.~Repko, and W.B.~Rolnick, \emph{The effective $W^\pm,
Z^0$ approximation for high energy collisions}, Phys. Lett. B
\textbf{148}, 367 (1984).
%
\bibitem{Cahn:1984}
R.N.~Cahn and S.~Dawson, \emph{Production of very massive Higgs
bosons}, Phys. Lett. B \textbf{136}, 196 (1984) [Erratum-ibid. B
\textbf{138}, 464 (1984)].
%
\bibitem{Cahn:1985}
R.N.~Cahn, \emph{Production of heavy Higgs bosons: comparisons of
exact and approximate results}, Nucl. Phys. B \textbf{255}, 341
(1985) [Erratum-ibid. B \textbf{262}, 744 (1985)].
%
\bibitem{Lindfors:1985}
J.~Lindfors, \emph{Distribution functions for heavy vector bosons
inside colliding particle beams}, Z. Phys. C \textbf{28}, 427
(1985).
%
\bibitem{Gunion:1986}
J.F.~Gunion, J.~Kalinowski, and A.~Tofighi-Niaki, \emph{Exact
calculation of $ff\rightarrow ffWW$ for the charged-current sector
and comparison with the effective-$W$ approximation}, Phys. Rev.
Lett. \textbf{57}, 2351 (1986).
%
\bibitem{Altarelli:1987}
G.~Altarelli, B.~Mele, and F.~Pitolli, \emph{Heavy Higgs production
at future colliders}, Nucl. Phys. B \textbf{287}, 205 (1987).
%
\bibitem{Lindfors:1987}
J.~Lindfors, \emph{Luminosity functions for $W^\pm$ and $Z^0$
initiated processes}, Z. Phys. C \textbf{35}, 355 (1987).
%
\bibitem{Johnson:1987}
P.W.~Johnson, F.I.~Olness, and W.-K.~Tung,
\emph{Effective-vector-boson method for high-energy collisions}
Phys. Rev. D \textbf{36}, 291 (1987).
%
\bibitem{Kuss:1996}
I.~Kuss and H.~Spiesberger, \emph{Luminosities for
vector-boson-vector-boson scattering at high energy colliders},
Phys. Rev. D \textbf{53}, 6078 (1996) [arXiv:hep-ph/9507204].
%
%
\bibitem{Ruiz:2021}
R.~Ruiz, A.~Costantini, F.~Maltoni, and O.~Mattelaer, \emph{The
effective vector boson approximation in high-energy muon
collisions}, JHEP \textbf{06}, 114 (2022) [arXiv:2111.02442].
%
\bibitem{Gupta:2012}
R.S.~Gupta, \emph{Probing quartic gauge boson couplings using
diffractive photon fusion at the LHC}, Phys. Rev. D \textbf{85},
014006 (2012) [arXiv:1111.3354].
%
\bibitem{Jikia:1994}
G.~Jikia and A.~Tkabladze, \emph{Photon-photon scattering at the
photon linear collider}, Phys. Lett. B \textbf{323}, 453 (1994)
[arXiv:hep-ph/9312228].
%
\bibitem{Gounaris:1999_1}
G.J.~Gounaris, P.I.~Porfyriadis, and F.M.~Renard,
\emph{Light-by-light scattering at high energy: a tool to reveal bew
particles}, Phys. Lett. B \textbf{323}, 76 (1994)
[arXiv:hep-ph/9812378].
%
\bibitem{Gounaris:1999_2}
G.J.~Gounaris, P.I.~Porfyriadis, and F.M.~Renard, \emph{The
$\gamma\gamma \rightarrow \gamma\gamma$ process in the standard and
SUSY models at high energies}, Eur. Phys. J. C \textbf{9}, 673
(1999) [arXiv:hep-ph/9902230].
%
\bibitem{Atag:2010}
S.~Ata\u{g}, S.C.~\.{I}nan and \.{I}.~\c{S}ahin, \emph{Extra
dimensions in $\gamma\gamma\rightarrow\gamma\gamma$ process at the
CERN-LHC}, JHEP \textbf{09}, 042 (2010) [arXiv:1005.4792].
%
\bibitem{Gounaris:1999_3}
G.J.~Gounaris, J.~Layssac, P.I.~Porfyriadis, and F.M.~Renard,
\emph{The $\gamma\gamma\rightarrow\gamma Z$ process at high energies
and the search for virtual SUSY effects}, Eur. Phys. J. C
\textbf{10}, 499 (1999) [arXiv:hep-ph/9904450].
%
\bibitem{Gounaris:2000}
G.J.~Gounaris, J.~Layssac, P.I.~Porfyriadis, and F.M.~Renard,
\emph{The $\gamma\gamma\rightarrow ZZ$ process and the search for
virtual SUSY effects at a $\gamma\gamma$ collider}, Eur. Phys. J. C
\textbf{13}, 79 (2000) [arXiv:hep-ph/9909243].
%
\bibitem{I_K:ampl}
S.C.~Inan, and A.V.~Kisselev, \emph{Probe of the
Randall-Sundrum-like model with the small curvature via
light-by-light scattering at the LHC}, Phys. Rev. D \textbf{100}
095004 (2019) [arXiv:1902.08615].
%
\bibitem{Berger:2005}
M.S.~Berger and B.~Zerbe, \emph{Signals for low scale gravity in the
process $\gamma\gamma\rightarrow ZZ$}, Phys. Rev. D \textbf{72},
095007 (2005) [arXiv:hep-ph/0507289].
%
\bibitem{SS}
G.~Cowan, K.~Cranmer, E.~Gross, and O.~Vitells, \emph{Asymptotic
formulae for likelihood-based tests of new physics}, Eur. Phys. J. C
\textbf{71}, 1554 (2011) [Erratum \emph{ibid}. \textbf{73}, 2501
(2013)] [arXiv:1007.1727].
%
\bibitem{Jacob:2000}
M.~Jacob and G.C.~Wick, \emph{On the general theory of collisions
for particlas with spin}, Ann. Phys. (N.Y.) \textbf{7}, 404 (1959);
Ann. Phys. \textbf{281}, 774 (2000).
%
\bibitem{Wigner}
E.P.~Wigner, \emph{Group theory and its application to the quantum
mechanics of atomic spectra}, Academic Press, New York, 1959.
%
\bibitem{Bateman_vol_1}
\emph{Higher Transcendental Functions}. Vol.~1. By the staff of the
Bateman manuscript project (A.~Erd\'{e}lyi,\emph{ Editor};
W.~Magnus, F.~Oberhettinger, F.G.~Tricomi, \emph{Associates}),
McGraw-Hill Book Company, New York, 1953.
%
\bibitem{Prudnikov_v3}
A.P.~Prudnikov, Yu.A.~Brychkov, and O.I.~Marichev, \emph{Integrals
and series}. Vol.~3: \emph{more special functions}, Gordon \& Breach
Sci. Publ., New York, 1989.
%
\bibitem{Randall:1999}
L.~Randall and R.~Sundrum, \emph{Large mass hierarchy from a small
extra dimension}, Phys. Rev. Lett. \textbf{83}, 3370 (1999)
[arXiv:hep-ph/9905221].
%
%
\bibitem{Fichet:2017}
S.~Fichet and C.~Baldenegro, \emph{Anomalous gauge interactions in
photon collisions at the LHC and FCC}, arXiv:1708.07531.
%
\bibitem{CMS-TOTEM:diphoton}
A.~Tumasyan \emph{et al.} (CMS and TOTEM collaborations),
\emph{First search for exclusive diphoton production at high mass
with tagged protons in proton-proton collisions at $\sqrt{s} = 13$
TeV}, Phys. Rev. Lett. \textbf{129}, 011801 (2022)
[arXiv:2110.05916].
%
\end{thebibliography}
\end{document}